\documentclass{iopart}
\usepackage{graphicx}
\newcommand{\be}{\begin{equation}}
\newcommand{\ee}{\end{equation}}
\renewcommand{\tr}{{\rm Tr}}
\newcommand{\Wg}{{\rm Wg}}
\renewcommand{\S}{\mathcal{S}}
\newcommand{\T}{\mathcal{T}}

\begin{document}

\title{A semiclassical matrix model for quantum chaotic transport}
\author{Marcel Novaes}
\address{Departamento de F\'isica, Universidade Federal de S\~ao Carlos, S\~ao Carlos, SP,
13565-905, Brazil}
\ead{marcel.novaes@gmail.com}

\begin{abstract}We propose a matrix model which embodies the semiclassical approach to the
problem of quantum transport in chaotic systems. Specifically, a matrix integral is
presented whose perturbative expansion satisfies precisely the semiclassical diagrammatic
rules for the calculation of general counting statistics. Evaluating it exactly, we show
that it agrees with corresponding predictions from random matrix theory. This uncovers
the algebraic structure behind the equivalence between these two approaches, and opens
the way for further semiclassical calculations.\end{abstract}


\maketitle

\section{Introduction}

It is well known that, in some regimes, random matrix theory (RMT) offers a good
description of properties of quantum systems whose classical limit displays chaotic
dynamics \cite{haake}. In particular, RMT has been successfully applied to the
calculation of several quantities in the context of electron transport \cite{beenakker},
in agreement with numerical simulations and experimental results involving, for example,
semiconductor quantum dots \cite{book}. The success of RMT resides in its
phenomenological character: instead of trying to predict the behavior of a specific
system, it treats the quantum $\S$-matrix as a random variable, and computes the average
value of observables in an ensemble of systems. The results are universal, in the sense
that they describe generic chaotic systems. The only input is the universality class,
corresponding to presence or absence of time-reversal invariance.

We have in mind a chaotic cavity, coupled to two ideal leads supporting respectively
$N_1$ and $N_2$ open channels. Transport can be characterized by a unitary $\S$-matrix of
dimension $N_1+N_2$, or by a $N_2\times N_1$ transmission matrix $t$. The quantities
${\rm Tr}[\T^n]$ are called transport moments or linear statistics, where $\T=t^\dag t$
(the first two such moments are related to the conductance and the shot-noise power).
They can be generalized to so-called non-linear statistics which are related, for
example, to conductance fluctuations. Within RMT, the $\S$-matrix is taken to be a random
unitary matrix from one of Dyson's circular ensembles \cite{beenakker} (we do not
consider the more general Poisson kernel \cite{poisson}, used when direct processes are
important).

A natural question is how to derive the universal RMT predictions from a semiclassical
description in terms of classical trajectories \cite{c3hub1993}, which is in principle
applicable to specific systems. It is now understood that the main ingredient is
constructive interference among trajectories that are almost identical (up to
time-reversal), except at small regions that are called `encounters' \cite{espalha}. This
permitted a diagrammatic formulation of the semiclassical theory, perturbative in the
parameter $(N_1+N_2)^{-1}$, that reproduced RMT results in a variety of situations
\cite{espalha,lead1,prl96sh2006,shot,njp9sm2007,jpa41gb2008,time,njp13gb2011,epl} (it is
even possible to go beyond RMT and take into account corrections due to the Ehrenfest
time, as discussed for example in \cite{Ehren}).

All semiclassical works up to now have relied on explicit and lengthy calculations based
either on group-theoretical manipulations involving factorizations of permutations, or on
graph-theoretical manipulations involving trees and other diagrams (the state of the art
can be found in \cite{GregJack}). These combinatorial aspects are interesting and have
revealed connections to unsuspected areas. However, a more direct demonstration of the
RMT-semiclassics equivalence would be desirable. In the present work we offer such a
demonstration, which uncovers the underlying algebraic structure behind that equivalence.
For simplicity, we focus here on systems with broken time reversal symmetry. More
detailed calculations and the treatment of other symmetry classes will be presented
elsewhere \cite{else}.

We proceed in two steps. First, we postulate a matrix integral which can be expressed
diagrammatically, by means of Wick's rule, with exactly the same diagrammatic rules that
govern the semiclassical approach. Second, we solve this integral, using the Andr\'eief
identity and the machinery of Weingarten functions, and show that it agrees with the RMT
prediction.

\section{Semiclassical Diagrammatics}
A non-increasing sequence of positive integers $\lambda_1,\lambda_2,\ldots$ is called a
partition of $n$ if $\sum_i\lambda_i=n$ and this is denoted by $\lambda\vdash n$. The
number of parts in $\lambda$ is $\ell(\lambda)$. The functions \be\label{nonlinear}
p_\lambda(\T)=\prod_{i=1}^{\ell(\lambda)}\tr(\T^{\lambda_i})\ee can be used to expand a
generic statistic. For systems with broken time-reversal symmetry, RMT predicts
\cite{prb78mn2008} \be\label{plambda} \langle
p_\lambda(\T)\rangle=\frac{1}{n!}\sum_{\mu\vdash n}\frac{[N_1]_\mu[N_2]_\mu}
{[N_1+N_2]_\mu}\chi_\mu(1^n)\chi_\mu(\lambda),\ee where the brackets denote an ensemble
average, $1^n$ represents the partition with $n$ unit parts, $\chi$ are the characters of
the permutation group $S_n$, and \be
[N]_\mu=\prod_{i=1}^{\ell(\mu)}\frac{(N+\mu_i-i)!}{(N-i)!}.\ee

In the semiclassical limit, the element $t_{oi}$ of the transmission matrix may be
approximated \cite{c3hub1993} by a sum over trajectories starting at incoming channel $i$
and ending at outgoing channel $o$. When we write a trace like Tr$\T^n$, we end up with
$n$ trajectories associated with the $t$'s (direct trajectories) and another set of $n$
trajectories coming from the $t^\dag$'s (partner trajectories). The structure of the
trace imposes that the direct ones take $i_k$ to $o_k$ (labels must be equal), while the
partner ones take $i_k$ to $o_{k+1}$ (labels are cyclically permuted). If we have a
product of traces, to each trace will correspond a cycle in the label permutation (for
details, see \cite{njp13gb2011}).

\begin{figure}
\includegraphics[scale=0.85,clip]{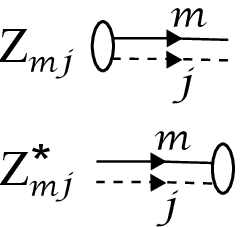}\hspace{1cm}\includegraphics[scale=1.,clip]{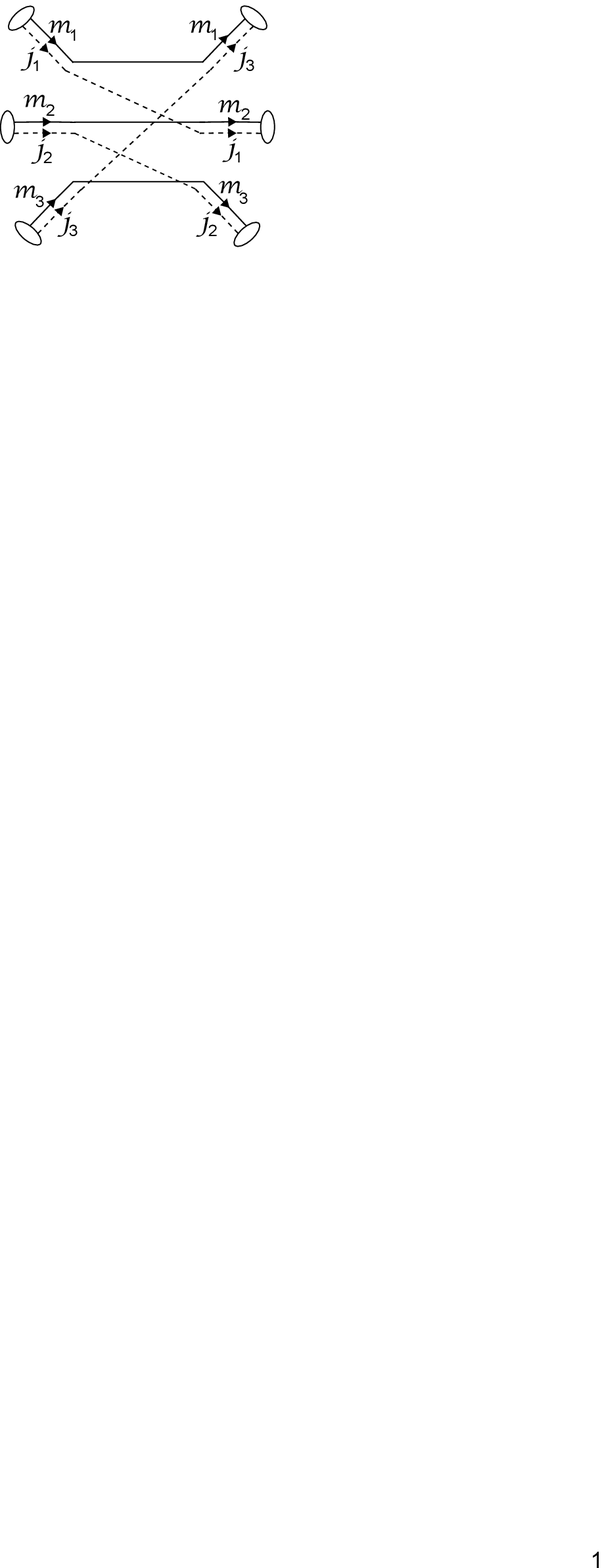}
\caption{Left: Diagrammatical representation of matrix elements of $Z$ and $Z^*$. The
former either arrives at a vertex or leaves the system, while the latter either emerges
from a vertex or enters the system. Right: A trace as a vertex with an internal
structure. Here we show
Tr$(ZZ^\dag)^3=Z_{m_1j_1}Z^\dag_{j_1m_2}Z_{m_2j_2}Z^\dag_{j_2m_3}Z_{m_3j_3}Z^\dag_{j_3m_1}$
(all indices summed over). }
\end{figure}

Systematic constructive interference survives an energy average only when each dashed
trajectory follows closely a solid one for a period of time, and they are allowed to
exchange partners at what is called an encounter. The two sets of trajectories are thus
nearly equal, differing only in the negligible encounter regions. These trajectory
multiplets are represented by a diagram whose edges are the regions where the
trajectories (almost) coincide and the internal vertices are the encounters. There are
$n$ external vertices of valence one on the left and on the right, corresponding to the
incoming and outgoing channels, respectively. The diagrammatic rule is that each internal
vertex produces $-(N_1+N_2)$, while each edge produces $(N_1+N_2)^{-1}$. This has been
discussed in detail in a number of papers \cite{prl96sh2006,shot,njp9sm2007,njp13gb2011}.

We shall define a matrix integral that mimics these diagrammatic rules. For that, we rely
on Wick's rule.

Suppose the matrix integral \be \langle f(Z,Z^\dag)\rangle=\frac{1}{\mathcal{Z}}\int dZ
e^{-M\tr ZZ^\dag}f(Z,Z^\dag),\ee where $\mathcal{Z}$ is a normalization factor, $M$ is a
parameter and $dZ$ denotes the flat measure on the space of $N$-dimensional matrices,
such that each matrix element is independently integrated over the whole complex plane.
Notice that $N$ is not related to channel numbers; its use is traditional in matrix
integrals. Instead, we shall later set $M=N_1+N_2$. Because of the Gaussian term, we
clearly have \be\label{cov} \langle
Z_{mj}Z^\dag_{qr}\rangle=\frac{\delta_{mr}\delta_{jq}}{M}.\ee Wick's rule,
\begin{eqnarray}\label{wick1}\fl \left\langle\prod_{k=1}^n
Z_{m_kj_k}Z^\dag_{q_{k}r_k}\right\rangle=\sum_{\sigma\in S_n}\prod_{k=1}^n\langle
Z_{m_kj_k}Z^\dag_{q_{\sigma(k)}r_{\sigma(k)}}\rangle=\frac{1}{M^n}\sum_{\sigma\in S_n}
\prod_{k=1}^n\delta_{m_kr_{\sigma(k)}}\delta_{j_{k}q_{\sigma(k)}},\end{eqnarray} is a
well known result \cite{zvon} . In words, it says we must sum, over all possible pairings
between $Z$'s and $Z^\dag$'s, the product of the average values of the pairs.

Wick's rule has an interesting diagrammatical representation \cite{diag1,diag2,diag3}.
Each matrix element $Z_{mj}$ is represented as a pair of arrows, one associated with $m$
and the other with $j$, with a marked end at the tail. In the matrix elements of $Z^*$,
the marked end is the head (see Figure 1). When computing an average like $\langle
p_\lambda(ZZ^\dag)\rangle$, we shall represent a term like $\tr(ZZ^\dag)^q$ by a vertex
of valence $2q$ with a specific internal structure, such that incoming $m_k$ is followed
by outgoing $m_k$, and incoming $j_k$ is followed by outgoing $j_k$. The marked ends are
all outside the vertex. An example is shown in Figure 1. Lines associated with $m$ labels
are drawn solid, lines associated with $j$ labels are drawn dashed.

Our ``semiclassical matrix integral'' is \be\label{model} G=\frac{1}{\mathcal{Z}}\int dZ
e^{-M\sum_{q\ge 1} \frac{1}{q}\tr[(ZZ^\dag)^q]} \prod_{k=1}^n
Z_{i_ko_k}Z^\dag_{o_{\pi(k)}i_k}.\ee It depends on two sets of $n$ indices, $i$ and $o$,
which will latter be summed over. It also depends on a permutation $\pi$ which
corresponds to the label permutation, i.e. to the statistics we wish to compute.

\begin{figure}
\includegraphics[scale=1.4,clip]{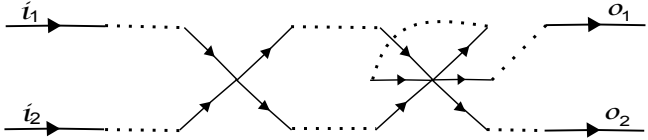}
\caption{Diagrammatical representation of one term coming from Wick's rule applied to the
average value of
Tr$(ZZ^\dag)^3$Tr$(ZZ^\dag)^2Z_{i_1o_1}Z^\dag_{o_2i_1}Z_{i_2o_2}Z^\dag_{o_1i_2}$. Dotted
segments are Wick connections. The permutation involved is $\pi=(12)$ and,
semiclassically, it contributes to $\langle {\rm Tr}\T^2\rangle$. We do not show dashed
lines and the internal structure of vertices, which appear in Figure 1. For clarity, only
the $i$-label of the incoming channels and the $o$-label of the outgoing ones are shown.}
\end{figure}

We take $e^{-M\tr ZZ^\dag}$ as part of the integration measure and expand the remaining
exponential as a power series in $M$. This produces all possible products of traces, each
trace being accompanied by a factor $-M$. Each trace is represented by a vertex as in
Figure 1. These are the internal vertices of the semiclassical diagrams. The matrix
elements $Z_{io}$ ($Z^\dag_{oi}$) are outgoing (incoming) arrows representing outgoing
(incoming) channels. These are the external vertices of the semiclassical diagrams.

We then integrate term by term using Wick's rule. All possible connections must be made
among vertices, using all marked ends of arrows, and each edge carries a factor $1/M$
from (\ref{cov}). The correct diagrammatic rule is therefore produced. We show an example
in Figure 2. The $\delta$-functions in (\ref{cov}) imply that we can associate to each
solid line the $i$ index of its incoming channel, and to each dashed line the $o$ index
of its outgoing channel.

The function $G$ will depend only on the cycle type of $\pi$, which we denote by
$\lambda$ (this means that the cycles of $\pi$ have lengths given by the parts of
$\lambda$). Linear statistics correspond to full cycles, having $\lambda=(n)$, while
non-linear statistics correspond to more general cycle structures. In fact, our matrix
integral is almost the same as $\langle p_\lambda(\T)\rangle$, where the brackets denote
an energy average. It is not exactly the same because, when we apply Wick's rule, some of
the resulting diagrams may contain periodic orbits, and this is not allowed in
semiclassical diagrams.

We can fix this problem with a little trick. Notice that the sum over the index
associated to the periodic orbit is free and produces a factor $N$. Hence, the
contribution of a diagram with $t$ periodic orbits will be proportional to $N^t$. In
order to exclude periodic orbits from our semiclassical theory, we simply have to
consider the part of the result that is independent of $N$. Equivalently, we may take the
limit $N\to 0$. This is possible since, once it is computed, $G$ is an analytic function
of $N$.

\section{Exact Solution}

In order to carry out the exact solution of our model (\ref{model}), we use some well
known facts about symmetric functions and characters that are summarized at the end of
the paper.

Let us start with the normalization constant $\mathcal{Z}=\int dZ e^{-M\tr ZZ^\dag}$. We
introduce the singular value decomposition \be\label{svd} Z=UDV^\dag,\quad DD^\dag=X={\rm
diag}(x_1,\ldots,x_N).\ee The measure $dZ$ can be written as \be
dZ=(\Delta(x))^2d\vec{x}dUdV,\ee where $\Delta(x)$ is the Vandermonde \be
\Delta(x)=\prod_{1\leq i<j\leq N}(x_j-x_i),\ee $d\vec{x}=dx_1\cdots dx_N$ and $dU$, $dV$
are Haar measures on the unitary group \cite{morris,Haar}. The angular integrals are
trivial; the remaining integral may be performed with the help of the Andr\'eief identity
(easily proved using the Leibniz formula for the determinant), \be \int
d\vec{x}\det(f_i(x_j))\det(g_i(x_j)) = N!\det \int dx f_i(x)g_j(x).\ee The result is that
\be\label{norm}\mathcal{Z}=\frac{N!}{M^{N^2}}\prod_{k=1}^{N-1}k!^2.\ee

The same change of variables turns (\ref{model}) into \be\label{model2} \int_0^\infty
\frac{d\vec{x}}{\mathcal{Z}}(\Delta(x))^2 \det(1-X)^M A_\pi(X),\ee where $A_\pi(X)=\int
dUdV a_\pi(X,U,V)$ and \be
a_\pi=\prod_{k=1}^n\sum_{j_k,m_k}U_{i_kj_k}D_{j_k}V^\dag_{j_ko_k}
V_{o_{\pi(k)}m_k}D^\dag_{m_k}U^\dag_{m_ki_k}.\ee We have written only one index for the
diagonal matrices, with an obvious meaning. The angular integrals may be expressed in
terms of so-called Weingarten functions $\Wg$ \cite{samuel,collins,esposti}. We have \be
A_\pi=\sum_{\sigma\tau\rho\theta\in
S_n}\Wg(\rho\theta^{-1})\Wg(\tau\sigma^{-1})\prod_{k=1}^n\sum_{j_k,m_k}B_kD_{j_k}D^\dag_{m_k},\ee
with \be B_k=
\delta_{i_ki_{\sigma(k)}}\delta_{j_km_{\tau(k)}}\delta_{o_{k}o_{\rho(\pi(k))}}
\delta_{m_kj_{\theta(k)}}.\ee

We now sum over $i$ and $o$, over the left and right channels, respectively, to get \be
\sum_{k=1}^n\sum_{i_k=1}^{N_1}\sum_{o_k=1}^{N_2}\delta_{i_ki_{\sigma(k)}}
\delta_{o_{k}o_{\rho(\pi(k))}}=N_1^{c(\sigma)}N_2^{c(\rho\pi)}=p_\sigma(1^{N_1})
p_{\rho\pi}(1^{N_2}),\ee where $c(\cdot)$ denotes the number of cycles of a permutation.
Also, since $\prod_k\delta_{j_km_{\tau(k)}}\delta_{m_kj_{\theta(k)}}=
\prod_{k}\delta_{j_km_{\tau(k)}}\delta_{m_km_{\tau\theta(k)}}$, the sum over the $j$'s is
simple, and the sum over the $m$'s gives \begin{eqnarray}
\sum_{k=1}^n\sum_{m_k}\delta_{m_km_{\tau\theta(k)}}x_{m_k}=p_{\tau\theta}(x).\end{eqnarray}
The Weingarten function depends only on the conjugacy class of its argument and has the
character expansion \be \Wg(\lambda)=\sum_{\mu\vdash n} w_\mu \chi_\mu(\lambda),\quad
w_\mu=\frac{1}{n!}\frac{\chi_\mu(1^n)}{[N]_\mu}.\ee Using the orthogonality relations for
characters then imply  \be\label{A2} A_\pi(x)=\sum_{\mu\vdash n}\frac{[N_1]_\mu[N_2]_\mu}
{([N]_\mu)^2}\chi_\mu(\pi)s_\mu(x),\ee where $s_\mu(x)$ is a Schur function.

Going back to (\ref{model2}), the integral over the eigenvalues is \be
E=\frac{1}{\mathcal{Z}}\int_0^1d\vec{x}(\Delta(x))^2 {\rm det}(1-X)^Ms_\mu(x).\ee The
determinantal form of the Schur function, together with the Andr\'eief identity, yields
\be E=\frac{N!M!^N}{\mathcal{Z}}{\rm
det}\left[\frac{(N+\mu_j-j+i-1)!}{(N+\mu_j-j+i+M)!}\right].\ee Standard manipulations
with determinants then lead to \be\label{E} E=\frac{M^{N^2}([N]_\mu)^2 \chi_\mu(1^n)}{n!}
\prod_{i=1}^N\frac{(M+N-i)!}{(\mu_i-i+2N+M)!}.\ee

Combining (\ref{E}) with (\ref{A2}), we get \be G=\frac{1}{n!} \sum_{\mu\vdash
n}[N_1]_\mu[N_2]_\mu\chi_\mu(1^n)\chi_\mu(\pi)g(N,M,\mu),\ee where \be
g(N,M,\mu)=M^{N^2}\prod_{i=1}^N\frac{(M+N-i)!}{(\mu_i-i+2N+M)!}.\ee The last step is the
limit $N\to 0$, which is quite simple: $\lim_{N\to 0}g(N,M,\mu)=1/[M]_\mu$. Identifying
$M$ with total number of channels, $M=N_1+N_2$, we finally obtain \be \lim_{N\to
0}G(\lambda)=\frac{1}{n!}\sum_{\mu\vdash n}\frac{[N_1]_\mu[N_2]_\mu}
{[N_1+N_2]_\mu}\chi_\mu(1^n)\chi_\mu(\lambda),\ee where again $\lambda$ is the cycle type
of $\pi$. Therefore, the RMT prediction (\ref{plambda}) is recovered.

\section{Conclusions}

We have introduced a new approach to the semiclassical calculation of quantum transport
properties in chaotic systems, which consists in building an auxiliary matrix integral
that has the correct diagrammatic rules and can be exactly solved. As an illustration, we
derived the counting statistics for broken time-reversal symmetry and arbitrary numbers
of channels. Counting statistics for other symmetry classes will be reported elsewhere
\cite{else}.

We believe this method will open the way to semiclassical calculations that were
previously unavailable, and may even provide results beyond RMT. For example, it may be
adapted to treat problems where the semiclassical trajectories have different energies,
as is necessary in calculations involving time delay \cite{time} or the proximity gap in
Andreev billiards \cite{Andreev}.

We also believe that similar ideas can be applied to closed systems, allowing the
calculation of all spectral correlation functions. This would provide justification for
the celebrated Bohigas-Giannoni-Schmit conjecture \cite{BGS} beyond the currently
understood form factor \cite{haakepre}. In fact, for broken time-reversal symmetry and
short times, this program has already been carried out \cite{Rn}.

\section*{Appendix}

For completeness, we present a summary of facts about symmetric functions and characters,
which have been used in this paper. They can be found in many standard references.

Let $x$ denote a set of $N$ variables. The power sum symmetric polynomials \be
p_\lambda(x)=\prod_{i=1}^{\ell(\lambda)} \sum_{j=1}^N x_j^{\lambda_i}\ee with
$\lambda\vdash n$ form a basis for the vector space of homogeneous symmetric polynomials
of degree $n$. It can be defined for matrix argument as in (\ref{nonlinear}). Another
basis for this space are Schur functions $s_\lambda(x)$, which can be written as a ratio
of determinants, \be\label{schur} s_\lambda(x)=\frac{{\rm det}(x_i^{\lambda_j-j+N})}{{\rm
det}(x_i^{j-1})}=\frac{{\rm det}(x_i^{\lambda_j-j+N})}{\Delta(x)}.\ee It is known that if
all $N$ variables in a Schur function are equal to $1$, it reduces to \be\label{1N}
s_\lambda(1^N)=\prod_{1\leq i<j\leq
N}\frac{\lambda_i-\lambda_j-i+j}{j-i}=\frac{\chi_\lambda(1^n)[N]_\lambda}{n!}.\ee

Let $S_n$ be the group of all $n!$ permutations of $n$ symbols. The above families of
functions are related by \be\label{p2s} p_\lambda(x)=\sum_{\mu\vdash
n}\chi_\mu(\lambda)s_\mu(x),\quad s_\lambda(x)=\sum_{\mu\vdash
n}\frac{\chi_\lambda(\mu)}{z_\mu}p_\mu(x)\label{s2p}, \ee where $z_\lambda$ is the order
of the centralizer of the conjugacy class in $S_n$ containing all permutations of cycle
type $\lambda$. Relation (\ref{s2p}) can also be written as a sum over permutations as
\be s_\lambda(x)=\frac{1}{n!}\sum_{\pi \in S_n}\chi_\lambda(\pi)p_\pi(x).\ee

The quantity $\chi_\mu(\pi)$ is the character of permutation $\pi$ in the irreducible
representation labeled by partition $\mu$. It depends only on the conjugacy class of
$\pi$. These quantities satisfy the orthogonality relations \be \sum_{\mu\vdash
n}\chi_\mu(\lambda)\chi_\mu(\omega)=z_\lambda\delta_{\lambda,\omega}, \quad
\sum_{\lambda\vdash n}\frac{1}{z_\lambda} \chi_\mu(\lambda) \chi_\omega(\lambda)=
\delta_{\mu,\omega}.\ee The latter is generalized as a sum over permutations as \be
\sum_{\pi\in S_n}\chi_\mu(\pi) \chi_\omega(\pi\sigma)=
n!\delta_{\mu,\omega}\frac{\chi_\mu(\sigma)}{\chi_\mu(1^n)}.\ee

\ack

I gratefully acknowledge interesting conversations with Gregory Berkolaiko and Sebastian
M\"uller, as well as financial support from CNPq and from grant 2012/00699-1, S\~ao Paulo
Research Foundation (FAPESP).


\begin{thebibliography}{99}
\bibitem{Haar} Actually $V$ is uniformly distributed in the coset space
    $\mathcal{U}(N)/[\mathcal{U}(1)]^N$, and $dV$ is the measure in this space induced
    from Haar measure in $\mathcal{U}(N)$. This difference is irrelevant here.
\end{thebibliography}
\end{document}